# Plutonium-Based Heavy-Fermion Systems

E. D. Bauer and J. D. Thompson*

Los Alamos National Laboratory

Los Alamos, NM 87545 USA

Abstract: An effective mass of charge carriers that is significantly larger that the mass of a free electron develops at low temperatures in certain lanthanide- and actinide-based metals, including those formed with plutonium, due to strong electron-electron interactions. This heavy-fermion mass is reflected in a substantially enhanced electronic coefficient of specific heat $\gamma$, which for elemental Pu itself is much larger than that of normal metals. By our definition, there are twelve Pu-based heavy-fermion compounds, most discovered recently, whose basic properties are known and discussed. Relative to other examples, these Pu-based heavy-fermion systems are particularly complex due in part to the possible simultaneous presence of multiple, nearly degenerate $5f^n$ configurations. This complexity poses significant opportunities as well as challenges, including understanding the origin of unconventional superconductivity in some of these materials.

email addresses:

E. D. Bauer, edbauer@lanl.gov

J. D. Thompson, jdt@lanl.gov

*corresponding author

J. D. Thompson, MS K764, Los Alamos National Laboratory, Los Alamos, NM 87545

1. INTRODUCTION

Plutonium (PU), element 94 in the periodic table, is a wonderfully complex metal whose structural and electronic properties are controlled by its 5f electrons. Though its nuclear properties are well-documented, the nature and consequences of its 5f electrons are not and this is reflected both in elemental Pu and compounds based on it.

1.1 Plutonium's 5f Electrons

Below its unusually low melting temperature of 913 K, Pu adopts six allotropic phases, with its least dense phase, $\delta$-Pu, having a negative thermal expansion coefficient (1). Though $\delta$-Pu exists from 592 K to 724 K in pristine form, it can be stabilized to low temperatures by small additions of group III or IV elements. From conventional electronic-structure calculations, $\delta$-Pu should be magnetic (2), but instead it is paramagnetic with a low temperature electronic specific heat that is larger by nearly an order of magnitude than all simple metals. Its allotropic cousin $\alpha$-Pu, although 20% denser, has a similarly enhanced electronic specific heat (1). These unusual characteristics of Pu are a consequence of its 5f electrons, which are, in a sense, indecisive—i.e., unable to choose to be localized close to the ionic core or to be delocalized and contribute to the Fermi sea – and so these electrons strike a compromise that minimizes their energy. When placed in certain crystal-chemical environments, one choice the 5f electrons make is to entangle themselves with conduction electrons to create itinerant quasiparticles with a heavy effective mass m*. We do not know in detail why or how heavy quasiparticles form in Pu-based metals, but a similar state emerges at low temperatures in other highly correlated metallic compounds with a periodic array of some lanthanide and actinide elements with a partially filled f-shell. Cerium, ytterbium, and uranium are notable examples of other f-elements that are prone to heavy-fermion behavior, and like Pu, their f-shell configuration is unstable. In isolation, the electronic configuration of Pu is [Radon]$5f^6 7s^2$, but the f-occupancy can vary from 2 to 6 when Pu is incorporated into a solid.

A few examples of Pu-based heavy-fermion systems have been known since the mid-1980's, but interest in their study has revived with the more recent discovery of unconventional superconductivity in $PuCoGa_5$ and related heavy-fermion materials. These so-called Pu115 compounds have received most attention and are providing a perspective on commonalities among classes of strongly correlated f-electron systems. Like offspring, we expect these Pu materials to inherit some parental characteristics, which are revealed both when Pu occurs as dilute impurities and when it is in concentrated form.

## 1.2 Lessons from the Parent

Dilute concentrations of Pu in a metallic host, such as La, appear to act as Kondo impurities (3). For Kondo-impurity behavior to exist, hybridization $V_{kf}$ of atomic-like f-orbitals with conduction-band electrons must be small relative to on-site Coulomb repulsion U. A small ratio of $|V_{kf}|^2/U$ leads to local f-moments that are magnetic (4), and these moments couple antiferromagnetically with spins of the electrons provided by the host through an exchange interaction $J \propto |V_{kf}|^2 U/E_f(E_f+U)$, where $E_f$ is the energy of the f-level relative to the Fermi energy $E_F$ (5). When U dominates, charge hybridization is negligible, and well below a characteristic (Kondo) temperature $T_K$, a virtual bound state emerges in which the f-moment and conduction-band spins are anti-aligned. As the singlet forms, a many-body resonance develops in the electronic density of states near $E_F$, with a resonance width $\Gamma_K = \pi N_0 V^2_{kf} \sim k_B T_K \propto 1/N_0[\exp(-$ $1/2JN_0)]$, where $N_0$ is the bare density of conduction states of the host. When the resonance is centered at the Fermi energy, the Sommerfeld coefficient of specific heat per impurity $\gamma_i \equiv C/T|_{T\to 0} \propto m^* \propto 1/T_K$.

At temperatures well above $T_K$, properties of compounds based on a dense array of certain f-elements, like Pu or Ce, resemble those expected of a collection of non-interacting Kondo impurities, and it has been reasonable to attribute their low-temperature, heavy-fermion state to the presence of a Kondo-like narrow resonance. Because of translational symmetry of the f-sublattice, however, an impurity model in not applicable, and an appropriate description must take the form of Bloch states of heavy quasiparticles that are more than a superposition of impurity resonances. A schematic illustration of the difference in the electronic spectrum for an impurity and lattice of impurities is shown in Figure 1.

From a Kondo-like perspective, $\delta$-Pu's electronic coefficient of specific heat $\gamma$ of 60 mJ/mol $K^2$ would imply a $T_K$ of several hundred Kelvin. This perspective, however, is not correct in detail. Recent dynamical mean field calculations for $\delta$-Pu find that its ground state is a quantum superposition of two valence configurations, $5f^5$ and $5f^6$, giving a time-average f-occupancy of 5.2 (6). Conclusions of these calculations are supported by resonant X-ray emission spectroscopy experiments (7). Such a non-integral occupancy is expected when the f-configuration fluctuates by exchanging electrons with the conduction-band sea, which is allowed if the hybridized 5f states are close to the Fermi energy, that is, when the ratio $|V_{kf}|^2/|E_F - E_f|$ becomes significant. In this case, an enhanced electronic specific heat, and hence m*, is due to a large electronic density of states that arises from the proximity of weakly dispersing hybridized f-bands to the Fermi energy, where the width of the hybridized bands $\Gamma_h \sim 1/\gamma$.

Hybridized bands are somewhat broader in the dense, low-symmetry $\alpha$-phase of plutonium (8), consistent with its smaller $\gamma$. Calculations similar to those performed on $\delta$-Pu show that $\alpha$-Pu is even more complex: Electronic correlations simultaneously allow multiple degrees of localization/delocalization at its eight inequivalent Pu sites (9).

Experimentally, it can be difficult to distinguish whether a Kondo-like or fluctuating-valence-like description is more appropriate, in part because spin fluctuations are inherent to the Kondo effect but charge fluctuations also carry spin fluctuations with them. A measure of the f-occupancy is useful and should be essentially integral in the Kondo-limit, but this determination is not always straightforward. When hybridization is weak compared to U, the Kondo scale typically is smaller than crystal-field splitting, as it is in some Ce-based heavy-fermion systems in which $\gamma$ can exceed 1000 mJ/mol $K^2$. In this limit, the temperature-dependent electrical resistivity frequently exhibits a Kondo-impurity-like signature of a minimum, followed at lower temperatures by a logarithmic increase. With further decreasing temperature, the resistivity drops toward zero as heavy Bloch states form. However, when the Kondo scale is large or the width of hybridized f-bands with fluctuating (mixed) valence is comparable to or greater than crystal-field splitting, the resistivity often is bulgy but decreases monotonically with decreasing temperature, which is the case for both $\alpha$- and $\delta$-Pu (1). In general, which scenario is appropriate depends on the crystal-chemical environment of the f-atom: This environment influences the strength of hybridization relative to U and $|E_F - E_f|$. Indeed, heavy-fermion behavior in Ce-based compounds spans a broad spectrum, from being in the Kondo limit and having a huge electronic specific heat with $\gamma$ of order 1000 mJ/mol $K^2$ to exhibiting signatures of a fluctuating f-configuration in compounds with $\gamma$ in the range of 100-300 mJ/mol $K^2$ (10, 11). Even though there is evidence for Kondo-impurity behavior in dilute Pu materials, there are so far no Pu-based heavy-fermion compounds with a huge $\gamma$. As we will discuss, Pu-based heavy-fermion materials are particularly enigmatic compared to other examples based on Ce, Yb, or even U and Np. In part, the challenge of studying Pu-heavy-fermion compounds arises from the multiple 5f-configurations that Pu hosts (12), ranging from $5f^2$ to $5f^6$, which stem from the more extended 5f wavefunction of Pu compared to the more localized 4f wavefunction of, for example $4f^5$ $Sm^{3+}$ (Figure 2), not to mention the strong spin-orbit coupling of actinides. Moreover, the substantial radioactivity of $^{239}Pu$, associated self-heating and self-damage, and significant neutron absorption present difficulties in conducting experiments on Pu-based materials.

## 2. Pu-BASED HEAVY-FERMION SYSYEMS

As implied in Section 1, heavy-fermion metals are identified experimentally by their large Sommerfeld coefficient of specific heat. In typical metals, $\gamma$ is a few mJ/mol $K^2$, which corresponds to an m* that is one to a few times the mass of a free electron. For our purposes, we arbitrarily define a heavy-fermion metal as one with a Sommerfeld coefficient of ~100 mJ/mol $K^2$ or greater.

### 2.1 Early Examples

Using this definition, $PuBe_{13}$, with $\gamma = 260 \pm 50$ mJ/mol $K^2$, was the first Pu-based heavy-fermion system identified by specific heat measurements (13). Above ~100 K, its magnetic susceptibility is Curie-Weiss-like with an effective moment of 0.74 $\mu_B$ (14), somewhat reduced from the Hund's-rule value of 0.84 $\mu_B$ for a localized $5f^5$ ($Pu^{3+}$) configuration, and the nonmonotonic temperature evolution of its electrical resistivity is typical of Kondo-lattice systems (13). Unlike its very heavy-fermion counterpart $UBe_{13}$ (15), $PuBe_{13}$ is not superconducting above 0.08 K but instead exhibits a broad maximum in the specific heat divided by temperature C/T, as displayed in Figure 3 that either could be Kondo-derived (13) or reflect a broadened antiferromagnetic transition (14). Very little else is known about the solid state properties of $PuBe_{13}$, in part because $^{239}PuBe_{13}$ is an intense neutron generator.

$PuAl_2$, one of the most studied heavy-fermion materials based on Pu, has a Sommerfeld coefficient of approximately 260 mJ/mol $K^2$ (16), which is similar to that of $PuBe_{13}$. Besides a large $\gamma$, specific heat measurements find anomalies at 3.5 and 9.5 K, with the lower temperature feature possibly reflecting a spin-density transition and the other possibly a structural transition, neither of which are revealed clearly in other measurements (16). The magnetic susceptibility of $PuAl_2$ follows a modified Curie-Weiss form, with an effective moment of 1.1 $\mu_B$, Weiss temperature of -150 K and temperature independent contribution of $1.5\text{x}10^{-4}$ emu/mol (17). This effective moment exceeds that for Hund's coupling in a $5f^5$ manifold, suggesting the coupling of total angular momentum ( j-j coupling) or intermediate coupling due to the strong spin-orbit coupling and/or an admixture of $5f^n$ states (18), but none of these possibilities has been verified. In contrast to a linear relation between the uniform susceptibility $\chi(0,0)$ and Knight shift (17), spin dynamics, reflected in the $^{27}Al$ spin relaxation rate $1/T_1$, scales as $T/\chi^2$ (0,0), which Fradin et al. (19) argued is a consequence of spin fluctuations in a 5f band strongly hybridized with the Al-derived s-band. With this interpretation, we might expect a bulgy electrical resistivity, but, instead, the temperature-dependent resistivity (17) is similar to that of $PuBe_{13}$. This apparent inconsistency may be understood if the 5f band, although strongly hybridized, is quite narrow, i.e., the 5f states are nearly localized, and consequently, the resistivity is dominated by scattering of electrons from spin fluctuations in this narrow band. Indeed, the low temperature resistivity of annealed $PuAl_2$ increases at low temperatures as $\rho(T) = \rho_0 + AT^2$, where $\rho_0 \approx 40\ \mu\Omega cm$ and $A \approx 1\ \mu\Omega cm/K^2$, giving a ratio $A/\gamma^2$ very close to that typical of Ce- and U-based heavy-fermion systems (20). In contrast, $\rho(T)$ of self-damaged $PuAl_2$ decreases as $\rho_0 (1-BT^2)$, with $\rho_0 \approx 190\ \mu\Omega cm$ (17). A resistivity that decreases quadratically with temperature from a large residual value is characteristic of incoherent scattering by a collection of non-interacting Kondo impurities. In effect, self-damage appears to destroy the highly correlated Bloch states of a very narrow 5f band.

Qualitatively, these two examples are not all that different than other heavy-fermion systems – exhibiting a crossover from impurity-like properties at high temperatures to a Bloch state of Landau-Fermi heavy quasiparticles at low temperatures that is susceptible to instabilities, such as magnetic order. Magnetic fluctuations, whether Kondo-derived or a consequence of a fluctuating valence, reveal themselves in nuclear magnetic resonance (NMR) and in their scattering of band electrons. Pu, however, introduces additional complexities posed by its nuclear instability as well as by its potential of exhibiting site-selective degrees of localization/delocalization (9).

## 2.2 Recent Examples

Before coming to a broader discussion in Section 3, we summarize the basic characteristics of more recently discovered Pu-based heavy-fermion systems. Table I lists known Pu-based metallic compounds with a Sommerfeld coefficient greater than $\sim 100$ mJ/mol $K^2$.

$PuIn_3$ is cubic and exhibits heavy-fermion behavior in the paramagnetic state before ordering antiferromagnetically below $T_N$ = 14.5 K (21). Recent nuclear magnetic resonance measurements suggest itinerant antiferromagnetism, possibly an incommensurate spin-density wave (22). A fit of the specific heat just above $T_N$ to the form $C/T = \gamma + \beta T^2$ yields $\gamma$ = 307 mJ/mol $K^2$, indicating a substantial effective mass enhancement of the itinerant charge carriers, and $\beta$ = 1.2 mJ/mol $K^4$, corresponding to a Debye temperature $\theta_D$ = 186 K (21). $PuIn_3$ is the only Pu-based material in which any quantum oscillations—reflecting extremal orbits of conduction electrons around the Fermi surface in a magnetic field—have been observed. These de Haas-van Alphen measurements deep in the antiferromagnetic state reveal a Fermi-surface pocket near the [111] direction with an effective mass enhancement $m^* = 5\ m_e$ (23). Though the predicted effective mass reaches 15 $m_e$ (24), the observed small m* is likely due to strong scattering caused by the radioactive decay of $^{239}Pu$, which limits the possibility of detecting all but the lightest mass orbits. It also is possible that measured m* underestimates the intrinsic mass renormalization because the internal magnetic field generated by long-range order can suppress many-body correlations. The electrical resistivity exhibits typical heavy-fermion behavior, in which $\rho(T)$ is weakly temperature-dependent at high temperatures then rapidly decreases below 50 K as the coherent Kondo lattice develops (21).

Depending on the synthesis conditions, $PuGa_3$ forms in two different crystal structures, either hexagonal or rhombohedral, which are made up of different stacking arrangements of Pu and Ga layers (25). Both allotropes order magnetically and show evidence for heavy-fermion behavior (25). In hexagonal $PuGa_3$, anomalies in specific heat and magnetic susceptibility indicate antiferromagnetic order at $T_N$ = 25 K. A fit to C/T data below 10 K yields $\gamma$ = 200

mJ/molK$^2$ and $\beta$ = 1.26 mJ/mol K$^4$, corresponding to a Debye temperature $\theta_D$ = 221 K. A similar enhancement of the effective mass is found in the rhombohedral $PuGa_3$, in which $\gamma$ = 100 mJ/mol K$^2$ in its ferromagnetic state below 22 K. The magnetic susceptibility is consistent with localized $Pu^{3+}$ above 100 K, where the susceptibility follows a modified Curie-Weiss law $\chi(T) = C/(T-\theta_{CW}) + \chi_0$, with an effective magnetic moment $\mu_{eff}$ = 0.80 $\mu_B$, as expected for $Pu^{3+}$, and $\theta_{CW}$ = +14 K. These $PuGa_3$ compounds emphasize the importance of crystal chemistry and its role in hybridization. Though both have the same chemical composition, their Sommerfeld coefficients appear to differ by a factor of two, and electrons order antiferromagnetically in one structure but ferromagnetically in the other.

Orthorhombic $Pu_2Ni_3Si_5$ exhibits ferromagnetic order at $T_C$ = 65 K, which is replaced by antiferromagnetic order at $T_N$ = 35 K (26). At low temperature, a fit to C/T data between 6 K < T < 10 K of the form $C/T = \gamma + aT^2$, expected for antiferromagnetic magnons, yields $\gamma$ = 85 mJ/mol Pu K$^2$. The magnetic susceptibility follows a modified Curie–Weiss law above 100 K yielding an effective magnetic moment $\mu_{eff}$ = 0.98 $\mu_B$, and a Curie–Weiss temperature $\theta_{CW}$ = +40 K. Magnetization measurements at 50 K provide evidence for weak ferromagnetism below $T_C$ = 65 K with a saturated magnetization $M_{sat}$ = 0.08 $\mu_B$/Pu atom. The electrical resistivity reflects transitions at $T_C$ and $T_N$ and above 100 K is weakly temperature dependent. Two magnetic transitions also are found at $T_{m1}$ = 37 K and $T_{m2}$ = 5 K in specific heat measurements of monoclinic $Pu_2Co_3Si_5$, but the nature of the magnetic order is unknown (26). A linear fit to the C/T data in a magnetic field of H = 6 T yields $\gamma$ = 95 mJ/mol Pu K$^2$.

There are only four Pu-based superconductors. All are members of the $PuMX_5$ (M=Co, Rh; X=Ga, In) family and crystalize in the $HoCoGa_5$ tetragonal structure just like their 'Ce115' heavy-fermion counterparts $CeMIn_5$ (M=Co, Rh, Ir) (21, 27-29). The electronic contribution to the specific heat, determined by subtracting a suitable lattice contribution, is plotted in Figure 4 as $C_{el}/T$ vs T for the four superconductors $PuCoGa_5$, $PuRhGa_5$, $PuCoIn_5$, and $PuRhIn_5$ (21, 30). Anomalies in $C_{el}/T$ reflect the superconducting transition temperatures $T_c$ = 18.5 K for $PuCoGa_5$, $T_c$ = 8.7 K for $PuRhGa_5$, and $T_c$ = 2.5 K for $PuCoIn_5$. Heavy-fermion superconductivity also is found in $PuRhIn_5$ at $T_c$ = 1.7 K from resistivity, susceptibility and NMR, but experimental constraints have prevented specific heat measurements below $T_c$ (21). The jump $\Delta C$ in specific at $T_c$ provides an estimate of the Sommerfeld coefficient through the BCS relation $\Delta C/\gamma T_c = 1.43$. The specific heat jumps for $PuCoIn_5$, $PuRhGa_5$, and $PuCoGa_5$ yield $\gamma \sim$ 150, 50, and 100 mJ/mol $K^2$, respectively. These estimated values, which assume weak electron-boson coupling, are roughly consistent with $C_{el}/T$ just above $T_c$ (Figure 4).

The upper critical field $H_{c2}$ versus T is shown in Figure 5 for $PuMX_5$, for magnetic field applied along the tetragonal a- and c-axis (21, 30, 32, 33). The initial slope of $H_{c2}(T)$, $dH_{c2}/dT|_{Tc}$, is proportional to $m^{*2} \bullet T_c$, and for H||a gives m* = 48 $m_e$, 41 $m_e$, 160 $m_e$, and 215 $m_e$ for $PuCoGa_5$, $PuRhGa_5$, $PuCoIn_5$, and $PuRhIn_5$, respectively. In general, the critical field of a singlet superconductor (e.g., s-wave, d-wave) is dictated by two competing responses of superconducting electrons to a magnetic field: Orbital pair breaking when the kinetic energy of the supercurrents around the normal core of quantized flux vortices exceeds the superconducting condensation energy, and spin or Pauli limiting due to the suppression of superconductivity when the polarization energy of spins of superconducting electrons exceeds the condensation energy of electron pairs. The upper critical field for H||c for both $PuCoIn_5$ [$H_{c2}^c(0)$ = 10 T] and $PuRhIn_5$ [$H_{c2}^c(0)$ = 7 T] is much smaller than expected ($0.7 \bullet T_c \bullet dH_{c2}/dT|_{Tc} \approx$ 21 and 16 T, respectively) for purely orbital limiting (34), indicating that Pauli limiting may play an important role in determining the value of $H_{c2}$, just as in the $CeMIn_5$ superconductors (35, 36). However, properties of $PuMGa_5$ superconductors, discussed below, suggest that a fluctuating valence may be more important for their description and perhaps relevant for the apparent lack of anisotropy in $H_{c2}(T)$ of $PuCoGa_5$ (30) and the unusual shape of $H_{c2}(T)$ of $PuRhGa_5$, which has been interpreted within a two-band model involving heavy and light quasiparticles (37).

$^{69}Ga$ and $^{115}In$ nuclear magnetic and quadrupole resonance (NQR) measurements on the $PuMX_5$ materials have been particularly useful for probing the symmetry of the superconducting order parameter. The Knight shift of $PuCoGa_5$ reveals that the spin susceptibility decreases in the superconducting state, consistent with spin-singlet pairing (38). Furthermore, a power-law temperature dependence of the spin-lattice relaxation rate $1/T_1 \propto T^3$ just below $T_c$ as displayed in Figure 6, along with lack of a Hebel-Slichter peak at $T_c$, provides strong evidence for d-wave superconductivity with line nodes in the superconducting gap (38). This assertion is in accord with tunneling spectroscopy studies of $PuCoGa_5$ that find a four-fold modulation of the superconducting gap within the ab-plane, consistent with a $d_{x^2-y^2}$ order parameter symmetry (39). The same superconducting order parameter symmetry has been determined from more detailed NMR and thermodynamic measurements of the $CeMIn_5$ (M=Co, Rh, Ir) superconductors (36, 40, 41). Spin-relaxation measurements on $PuRhGa_5$, $PuCoIn_5$, and $PuRhIn_5$ also find $1/T_1 \propto T^3$ just below $T_c$ (42, 43). For all of the $PuMX_5$ compounds (Fig. 6), fits of $1/T_1$ to a dirty d-wave model, which includes the effects of scattering from impurities arising from the radioactive decay of $^{239}Pu$, indicate that $2\Delta/k_BT_c$ = 5-8. Relative to

the weak-coupling d-wave ratio $2\Delta/k_BT_c = 4.28$, these conclusions suggest that superconductivity in these materials is in the strong-coupling limit. If so, then their Sommerfeld coefficient estimated earlier from the specific heat jump at $T_c$ may be slightly too large.

The normal state of $PuMX_5$ crystals reflects the influence of strongly correlated 5f electrons (21). For example, the electrical resistivity of $PuCoGa_5$ is weakly temperature dependent above 200 K and has an s-shaped curvature below 150 K, typical of mixed-valent systems. Below about 50 K, $\rho(T)$ assumes an unusual power-law T-dependence with $\rho \propto T^{4/3}$ (28) that also appears in $PuRhGa_5$ (29). In contrast, $\rho(T)$ of $PuCoIn_5$ and $PuRhIn_5$ at intermediate temperatures is more typical of a heavy-fermion compound with a narrow 5f-band, in which the resistivity is nearly T-independent above 150 K and passes over a maximum around 100 K before decreasing rapidly at lower temperatures (31). Below about 10-20 K, $\rho(T) \propto AT^n$ with n = 1 and 4/3 for $PuRhIn_5$ and $PuCoIn_5$, respectively, power-laws that are distinctly different from the $T^2$ dependence of heavy Landau-Fermi quasiparticles (31). Specifically, the T-linear resistivity of $PuRhIn_5$ (31) from 20 K to $T_c = 1.7$ K and a power-law form of $C/T = \gamma_0 - AT^{1/2}$ from 2 K $\leq T \leq$ 6 K (Figure 4 inset) are characteristic of systems close to an antiferromagnetic quantum-critical point (44, 45) and reminiscent of properties of the quantum-critical superconductors $CeCoIn_5$ (46) and $CeCu_2Si_2$ (47).

As with the resistivity, NMR/NQR studies of the normal state are difficult to interpret straightforwardly. As displayed in Figure 6 for $PuCoIn_5$, $1/T_1$ is essentially temperature independent above 60 K, signaling the presence of weakly coupled 5f magnetic moments (43). Between 10 to 40 K, Korringa-like relaxation, $1/T_1 \propto T$, appears and is followed at lower temperatures by a faster relaxation, possibly due to the presence of antiferromagnetic spin fluctuations that also could be the source of the $T^{4/3}$ dependence of $\rho(T)$. Though the uniform static magnetic susceptibility $\chi(0,0)$ of $PuCoIn_5$ has not been measured, it is known for $PuRhIn_5$ whose relaxation rate is similar in magnitude to that of $PuCoIn_5$. In $PuRhIn_5$, $\chi(0,0)$ above ~100 K follows a modified Curie-Weiss form with an effective moment $\mu_{eff} \approx 0.85\ \mu_B$ ($0.3\ \mu_B$) and Weiss temperature $\theta_P \approx$ -120 K (-7 K) for a magnetic field perpendicular (parallel) to the c-axis. The negative and relatively small Weiss temperatures are consistent with antiferromagnetic correlations/interactions. The origin of anisotropy in both $\mu_{eff}$ and $\theta_P$ remains an open question but, given the small Weiss temperature, conceivably could result from crystal fields. As seen in Figure 6, the relaxation rates in $PuMGa_5$ are similar and notably slower than in the In analogs. In contrast to $PuRhIn_5$, $\chi(0,0)$ is weakly anisotropic in $PuRhGa_5$, even though its $\mu_{eff} = 0.85\ \mu_B$ also is consistent with a $5f^5$ configuration, and is only weakly temperature dependent in $PuCoGa_5$ (48). The later suggests relatively broad 5f bands, possibly consistent with mixed-valent behavior. Nevertheless, $1/T_1$ below about 30 K in $PuCoGa_5$ (38) is clearly not Korringa-like.

$PuPt_2In_7$ belongs to the $Pu_mM_nIn_{3m+2n}$ tetragonal family of compounds (49) of which the $PuMX_5$ compounds are a subset. In this n=2, m=1 bi-layer variant of the 'Pu115s' (n=1, m=1), there is no evidence for a phase transition in specific heat above 2 K. In contrast, $CePt_2In_7$ orders antiferromagnetically at $T_N$ = 5 K at ambient pressure and becomes superconducting near $P_c$ = 3 GPa where the AFM order is suppressed (50, 51). A linear fit to C/T vs $T^2$ for $PuPt_2In_7$ yields $\gamma$ = 250 mJ/mol $K^2$ (49). The resistivity of $PuPt_2In_7$ shows a broad maximum at ~100 K followed by a rapid decrease in $\rho(T)$ as heavy quasiparticles emerge in relatively narrow hybridized bands.

## 3 DISCUSSION

With the existence of well-defined crystal-field levels (52, 53) and an effective moment expected for $Ce^{3+}$, there is no doubt that the 4f configuration in $CeMIn_5$ compounds is nearly integral above 100 K or so. Nevertheless, the fate of the 4f electron at low temperatures depends on the ligand ion, being hybridized sufficiently to contribute to the Fermi volume when M=Co and Ir and remaining localized in $CeRhIn_5$ (54, 55). The Pu115s are far more complex. Although the unit cell volume of $PuMIn_5$ materials is only a few percent smaller than that of their corresponding Ce115 cousin, the configuration of Pu's 5f electrons is non-integral. As shown in Figure 7, RXES measurements (56) reveal an admixture of $5f^4$ ($Pu^{4+}$), $5f^5$ ($Pu^{3+}$), and $5f^6$ ($Pu^{2+}$) configurations in $PuCoIn_5$ at room temperature: a dominant $5f^5$ component (77% weight) that is nearly degenerate with $5f^4$ (21%) and $5f^6$ (2%) configurations. $PuCoGa_5$ is similarly mixed-valent (Figure 7), but with a somewhat more even distribution in configurational weight: 62% of $5f^5$, 29% of $5f^4$, and 9% of $5f^6$. In both, the average valence is $\nu = 3.2$, despite the 28% larger unit cell volume and longer c-axis of $PuCoIn_5$ relative to $PuCoGa_5$. Cell volume apparently is not the deciding factor in determining the 5f configuration. A lesson learned from theoretical (57) and experimental (T. Williers, F. Strigari, Z. Hu, V. Sessi, N. B. Brooks, E. D. Bauer, J. L. Sarrao, J. D. Thompson, A. Tanaka, S. Wirth. L. H. Tjeng, A. Severing, unpublished) studies of the Ce115s is that f-hybridization with the out-of-plane In electrons is an important consideration, and it is reasonable to suggest that this is the source of the difference in 5f-configurational weights in $PuCoIn_5$ and $PuCoGa_5$. Irrespective of the specific origin of this difference, it is difficult to reconcile qualitative differences in uniform susceptibility and relaxation rates in $PuCoX_5$ systems when their average valence is the same.

All $CeMIn_5$ (M=Co, Rh, Ir) materials have large Sommerfeld coefficients, of order $\gamma$ = 400-1000 mJ/mol $K^2$, and are in the Kondo limit (27, 59). The characteristic energy scale for spin fluctuations in them is of order $T_K$ = 20 -50 K, deduced from $T_K \approx Rln2/\gamma$, where R is the gas constant (38, 59). Moreover, these materials are in close proximity to an antiferromagnetic quantum-critical point (27), accessed either by pressure ($CeRhIn_5$) or chemical substitution ($CeRh_{1-x}Ir_xIn_5$), and show non-Fermi liquid temperature dependences in thermodynamic, spin-relaxation and transport properties. In phonon-mediated superconductors, the superconducting transition temperature is given by $T_c = \theta_D \exp(-1/\lambda)$, where $\theta_D$ is the Debye temperature and $\lambda$ is the electron-phonon coupling constant. Theoretical models suggest a similar relation for heavy-fermion superconductors (60), $T_c = \Gamma_h \exp[-1/g]$, but now $\Gamma_h$ is a characteristic energy of spin or charge fluctuations mediating the superconductivity and g is a measure of electron-boson coupling. As discussed in Section 1.1, $\Gamma_h$ is inversely proportional to the Sommerfeld coefficient or equivalently the heavy-quasiparticle density of states which reflects the extent of hybridization. As seen in Figure 8, heavy-fermion superconductors, including the Pu115s, follow a common linear relationship between $T_c$ and $\Gamma_\gamma$ (38) for a reasonable range of g-values (~0.9 - 1.2). For the lower $T_c$ Ce-based superconductors, such as the Ce115s and $CeCu_2Si_2$ (61), a preponderance of data favors the proposal that unconventional superconductivity is mediated by spin fluctuations. Whether this also is the case for the $PuMX_5$ superconductors remains to be established, but the correlation in Figure 8 suggests that superconductivity in them has an unconventional origin, e.g., spin- and/or valence- fluctuation-mediated pairing. There is a precedent in $CeCu_2Si_2$ for two pressure-induced domes of superconductivity, a lower $T_c$ dome arising from spin fluctuations and a higher $T_c$ dome developing from valence fluctuations (47). Further experiments will be useful to establish if two mechanisms also might be active in $PuMX_5$ systems.

Although evidence suggests the dominance of spin and/or valence fluctuations in producing superconductivity, there are, of course, phonons in the $PuMX_5$ compounds. A hallmark of phonon-mediated superconductivity is a change in $T_c$ induced by a change in isotopic mass $m_i$ with $T_c \propto (m_i)^{-1/2}$ (62). Such an isotope effect produced by replacing $^{239}Pu$ with $^{242}Pu$ in $PuCoGa_5$ should reduce $T_c$ by 0.02% or equivalently 0.04 K. This effect has not been found in measurements of $T_c$ in $^{239}PuCoGa_5$ and $^{242}PuCoGa_5$, but in the unlikely possibility (63) that it were to exist, it could be masked easily by the large effect of self-damage that suppresses $T_c$ in the former compound at a rate of -0.2 K/month (28).

## 4. OUTLOOK

Except for $PuAl_2$ and $PuBe_{13}$, Pu-based heavy-fermion materials are rather new, most appearing in the past ten years, and relative to other Ce-, U- or Yb-based heavy-fermion systems are unstudied except for their most basic properties. Consequently, this short review has been primarily descriptive rather than insightful. From this perspective, there are many opportunities to further guide our understanding of these remarkably complex materials. Although we have discussed them in contexts familiar from what is known primarily about Ce-based heavy-fermion systems, almost certainly this is naïve, and, indeed, it is not obvious that historical notions of the Kondo effect and mixed valence are even appropriate. The near degeneracy of multiple $5f^n$ configurations in Pu-based heavy-fermion systems compounds their complexity. More examples and many more experiments are needed to make progress on this challenging problem. With only a few institutions world-wide able to make and measure Pu materials, this is far easier said than done. However, theory does not have this limitation, and this is where there is hope for more rapid progress. With a limited number of experiments reasonable, theory could be particularly valuable by suggesting critical, key (but achievable) measurements that would truly advance the field. As with other heavy-fermion systems, extending Pu measurements to the lowest temperatures possible would be useful. This requires reducing self-heating and self-damage to as low as possible by preparing samples with the much less active, but rare, $^{242}$Pu isotope. This isotope also is more amenable to neutron-scattering experiments. In this regard, it will be particularly interesting to search for a neutron-spin resonance in the superconducting state of, for example, $PuCoGa_5$. Such a resonance, found now in cuprates, iron-pnictides and $CeCoIn_5$ (64-67), should appear at about 8 meV.

Using our generous criteria of a Sommerfeld coefficient of about 100 mJ/mol $K^2$ to ‘define’ heavy, there are only 12 Pu-based examples. Nevertheless, among these 12, there are paramagnetic, antiferromagnetic, ferromagnetic, superconducting and possibly other ground states. The existence of multiple $5f^n$ configurations, with both an even and odd number of electrons, raises the possibility of degenerate, non-Kramers electronic states with a variety of multipolar (e.g., quadrupolar, octupolar) orderings and associated fluctuations that have yet to be explored (68). We have discussed only metallic systems, but Kondo-like hybridization also can produce an insulating state if the chemical potential falls in the hybridization gap (Figure 1b). This is the case for $SmB_6$, which theory predicts to be a ‘topological Kondo insulator’ (69) and experiments appear to support (70-72). With even stronger spin-orbit coupling in isostructural $PuB_6$, it also is predicted to have a topologically protected metallic surface state (73). Experiments similar to those made on $SmB_6$ are, in principle, possible for $PuB_6$, and it will be very interesting to see what those experiments show.

Perhaps one of the more outstanding issues is the pairing mechanism active in producing superconductivity in the Pu115s. Spin fluctuations are a leading possibility but valence fluctuations, with their associated charge and spin fluctuations, are another. NMR/NQR, neutron scattering and elastic constant measurements that respond to these fluctuations will be especially worthwhile to help resolve this issue.

Disclosure Statement

The authors are not aware of any affiliations, memberships, funding, or financial holdings that might be perceived as affecting the objectivity of this review.

Acknowledgements

We thank M. Janoschek, G. Koutroulakis, J. Lawrence, A. Mounce and H. Yasuoka for useful discussions, J. Mitchell and P. Tobash for sample preparation, and C. Booth for providing the data for Figure 7.  Work at Los Alamos National Laboratory was performed under the auspices of the U.S. Department of Energy, Office of Basic Energy Sciences, Division of Materials Sciences and Engineering.

**Table 1:** Properties of Pu-based heavy-fermion materials. PM: paramagnetic. AFM: Antiferromagnetic. FM: Ferromagnetic. SC: Superconducting.

| Compound | Crystal Structure | Magnetic Order | $\gamma$ (mJ/mol Pu-K$^2$) | Reference |
|---|---|---|---|---|
| $PuBe_{13}$ | Cubic | PM | 260 | (13) |
| $PuAl_2$ | Cubic | Unknown type $T_{m1}$ = 9.5 K, $T_{m2}$ = 3.5 K | 260 | (16) |
| $PuIn_3$ | Cubic | AFM $T_N$ = 14.5 K | 300 | (21) |
| $PuGa_3$ | Hexagonal | AFM $T_N$ = 25 K | 200 | (25) |
| $PuGa_3$ | Rhombohedral | FM $T_C$ = 22 K | 100 | (25) |
| $Pu_2Ni_3Si_5$ | Orthorhombic | FM / AFM $T_C$ = 65 K, $T_N$ = 35 K | 85 | (26) |
| $Pu_2Co_3Si_5$ | Monoclinic | Unknown type $T_{m1}$ = 37 K, $T_{m2}$ = 5 K | 95 | (26) |
| $PuCoGa_5$ | Tetragonal | PM (SC, $T_c$ =18.5 K) | 75-100 | (28) |
| $PuRhGa_5$ | Tetragonal | PM (SC, $T_c$ = 8.7 K) | 50 | (29) |
| $PuCoIn_5$ | Tetragonal | PM (SC, $T_c$ = 2.5 K) | 200 | (31) |
| $PuRhIn_5$ | Tetragonal | PM (SC, $T_c$ = 1.7 K) | 350 | (21) |
| $PuPt_2In_7$ | Tetragonal | PM | 250 | (49) |

Figure captions

Figure 1. Schematic density of states (DOS) versus energy for (a) Kondo impurity and (b) Kondo lattice. Note the new feature of a hybridization gap, here centered just above $E_F$, in a Kondo lattice.

Figure 2. Radial probability distribution of wavefunctions of (a) $Sm^{3+}$ and (b) $Pu^{3+}$ showing the greater extent of the $Pu^{3+}$ 5f wavefunction relative to the 4f wavefunction of $Sm^{3+}$. Adapted with permission from reference (74).

Figure 3. Specific heat divided by temperature C/T versus T for $CeBe_{13}$, $UBe_{13}$, and $PuBe_{13}$. $CeBe_{13}$ is strongly mixed valent (75). Inset: Superconducting transition of $UBe_{13}$ at $T_c$ = 0.8 K. Data from references (13, 15, 75).

Figure 4. Electronic specific heat, plotted as $C_{el}/T$ versus T, for $PuMGa_5$ (M=Co, Rh) and $PuCoIn_5$. (a) Electronic specific heat of $PuRhIn_5$ showing the increase of $C_{el}/T$ with decreasing temperature, indicating proximity to a quantum critical point. The solid curve is a fit to the data of the form $\gamma_0$-$AT^{1/2}$. (b) Crystal structure of $PuCoGa_5$ with Pu (purple), Co (light blue), and In (orange). Data for $PuCoGa_5$, $PuRhGa_5$, $PuCoIn_5$ and $PuRhIn_5$ are from references 21 and 30.

Figure 5. Upper critical field $H_{c2}$ for $PuMGa_5$ (M=Co, Rh) and $PuMIn_5$ (M=Co, Rh) for H||a-axis (red symbols) and H||c-axis. Data from references 21 and 30-32.

Figure 6. Spin-lattice relaxation rate $1/T_1$ versus T for the $PuMGa_5$ and $PuMIn_5$ (M = Co, Rh) heavy-fermion superconductors. The solid black line is a $T^3$ power-law. The red dash-dot line shows the Korringa relation for $PuCoIn_5$, and the black dashed line shows the T-independent

behavior of $1/T_1$, taken as a measure of the Kondo temperature. Data from references 38, 42 and 43).

Figure 7. Resonant X-ray Emission Spectroscopy of $PuCoGa_5$ and $PuCoIn_5$ (56). (a) RXES spectra of $PuCoGa_5$ at T=300 K. Each curve (normalized to its own peak height) displays the emission intensity for a cut at incident energy $E_I$ vs transfer energy $E_T = E_I - E_e$, where $E_e$ is the emission energy between the intermediate 3d and final 2p core states. Plotted in this way, the difference of the $5f^4$ and $5f^6$ spectral weights appear to be the most distinct between the two compounds, but all three configurations change appreciably. Fits to the data of the Kramers-Heisenberg formula yield the unoccupied density of states (upDOS fit) shown in panel b. Note the clear difference in the upDOS between $PuCoGa_5$ and $PuCoIn_5$ that results from the change in configurational weight of $5f^4$, $5f^5$, and $5f^6$; however, the average valence of the two superconductors is the same ($\upsilon$ = 3.2).

Figure 8. Superconducting transition temperature $T_c$ versus hybridization strength $\Gamma_h$, as determined by the inverse of the density of states, for heavy-fermion superconductors (38).

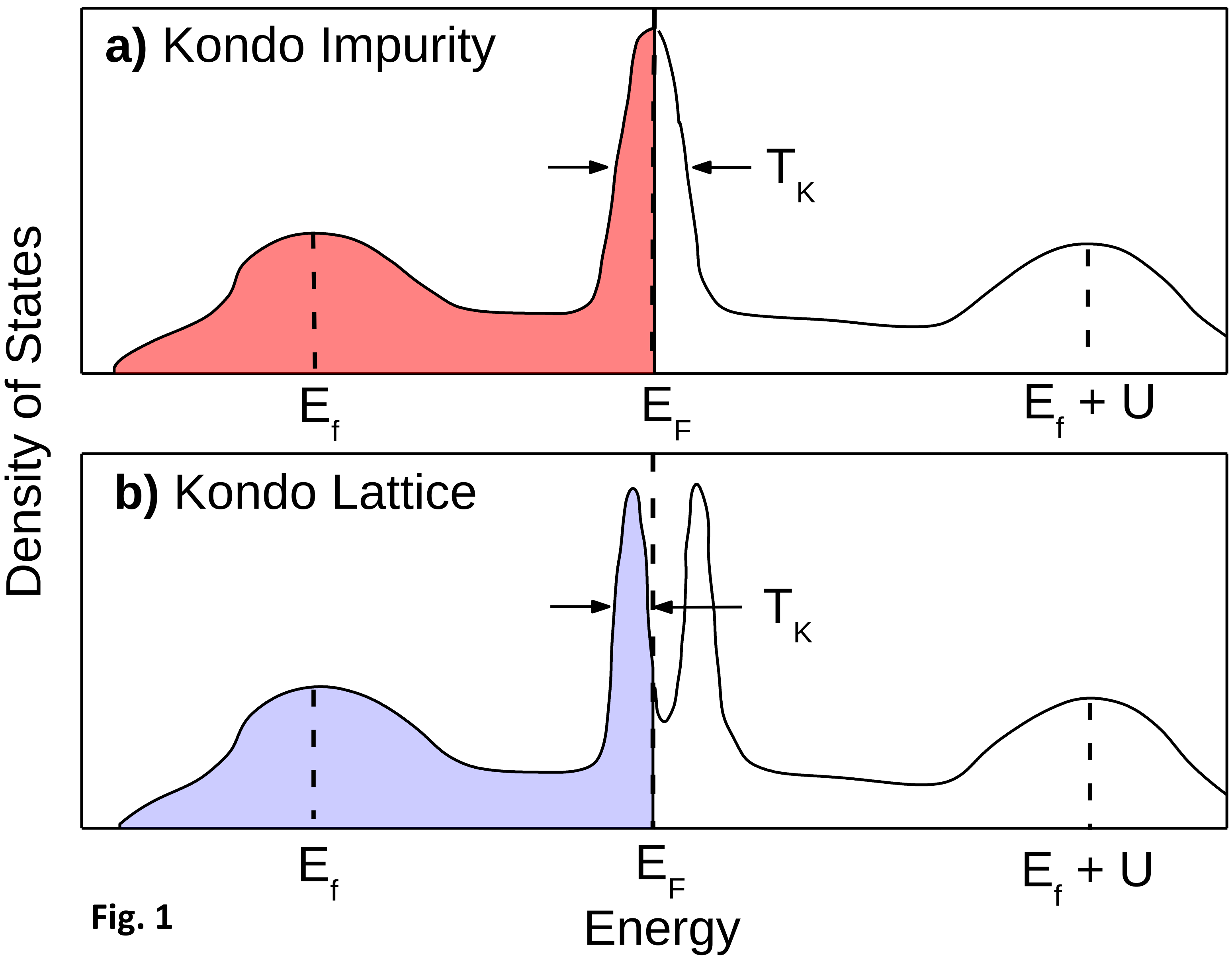

**Fig. 1**

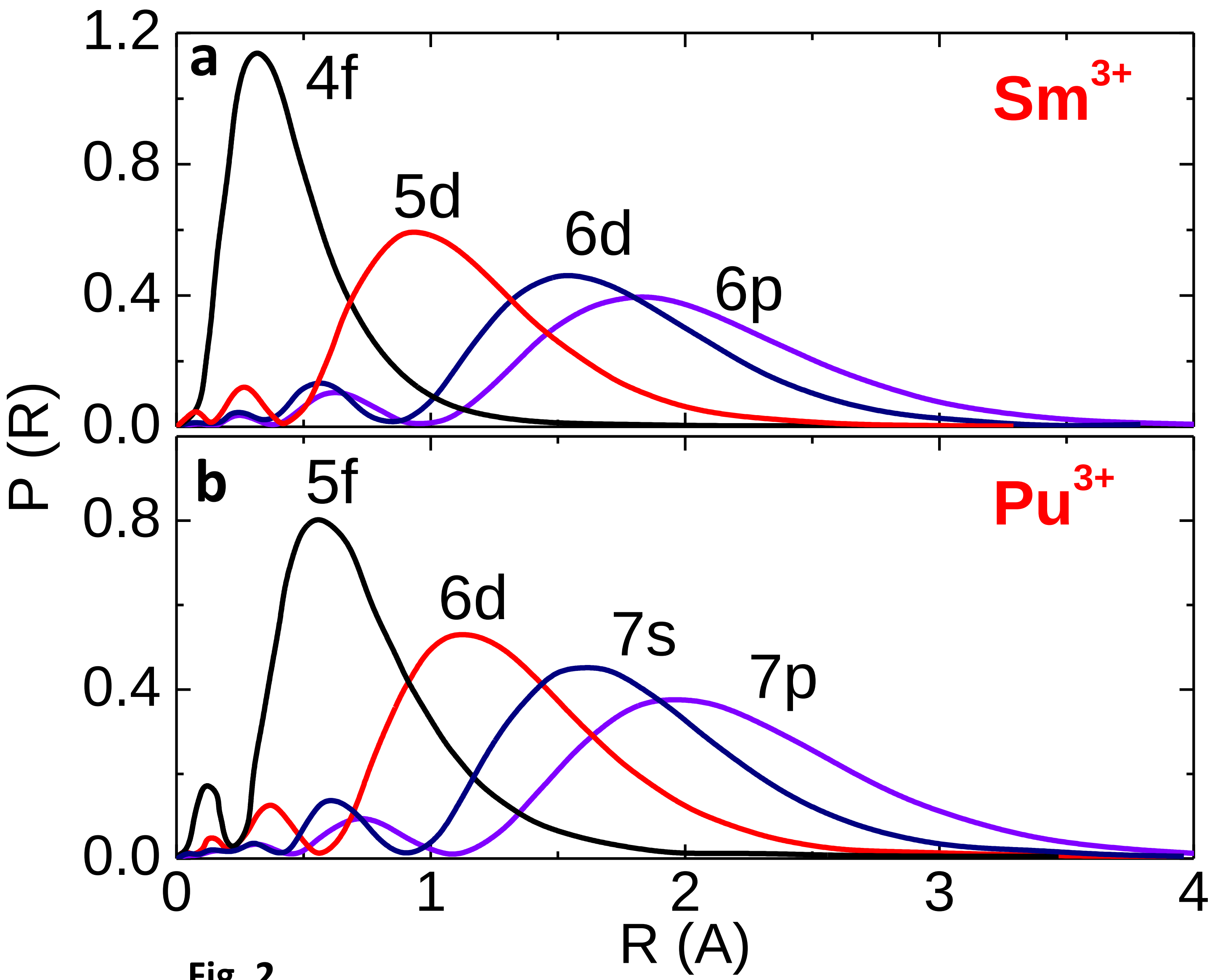

**Fig. 2**

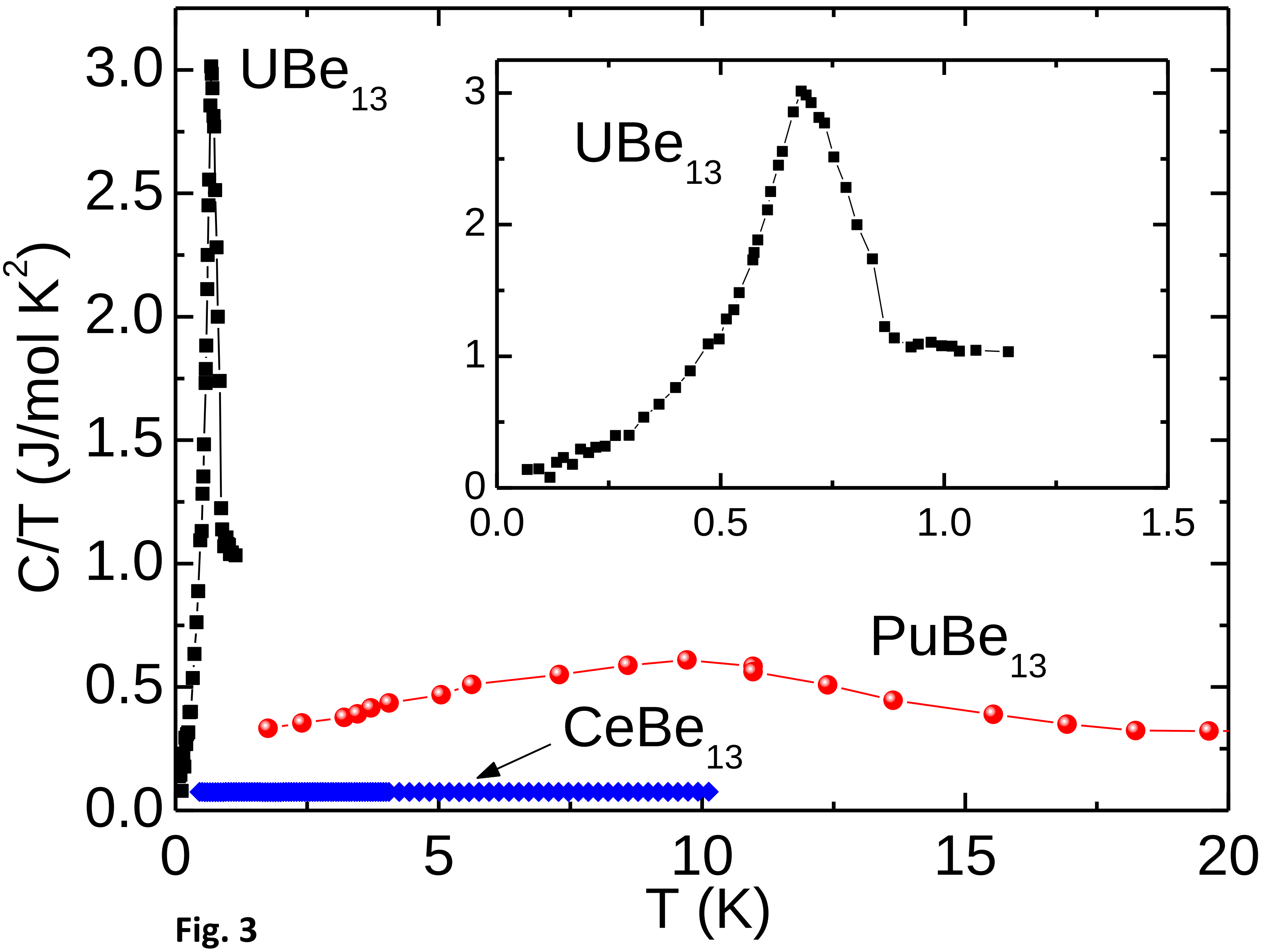

**Fig. 3**

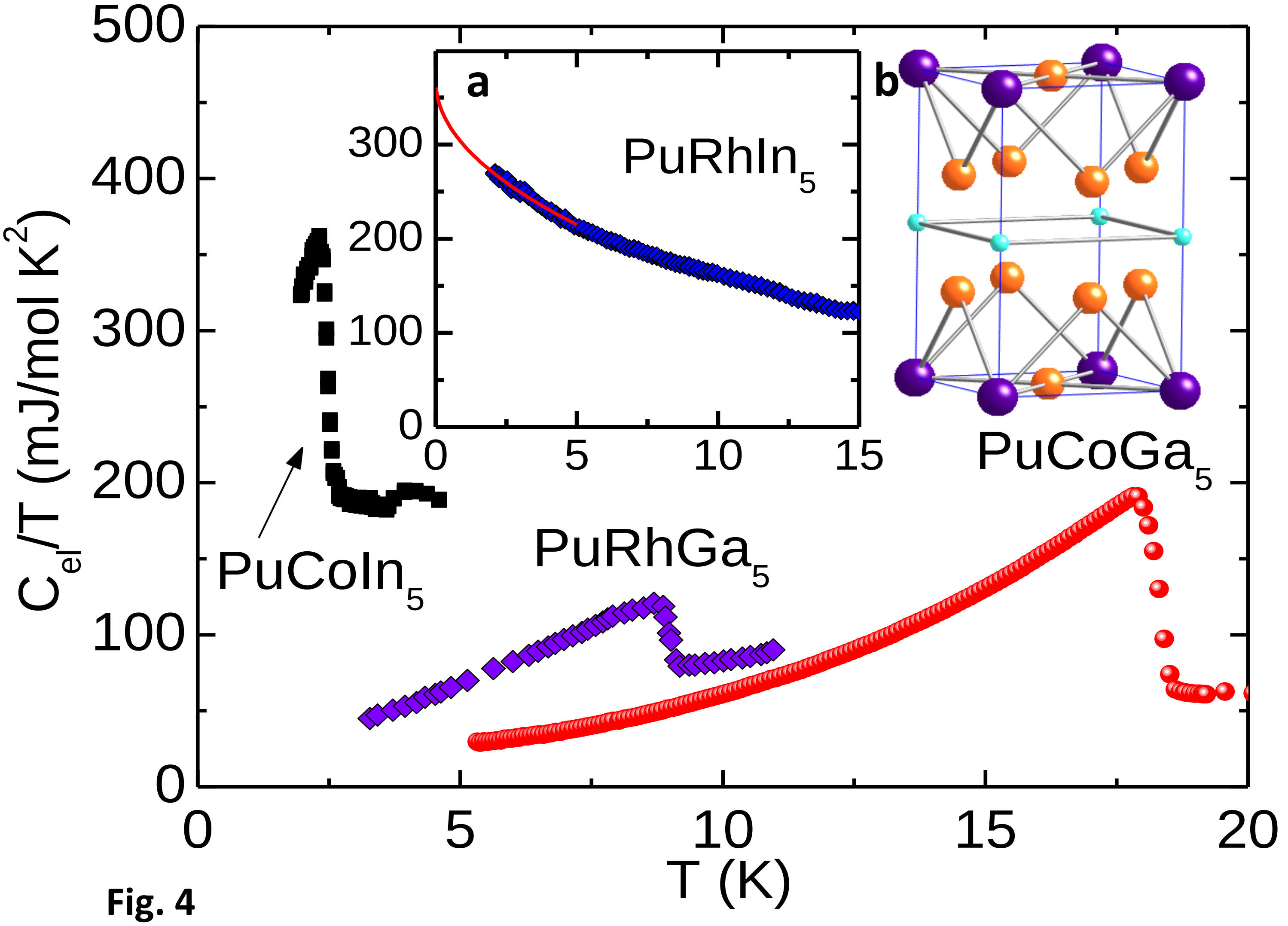

**Fig. 4**

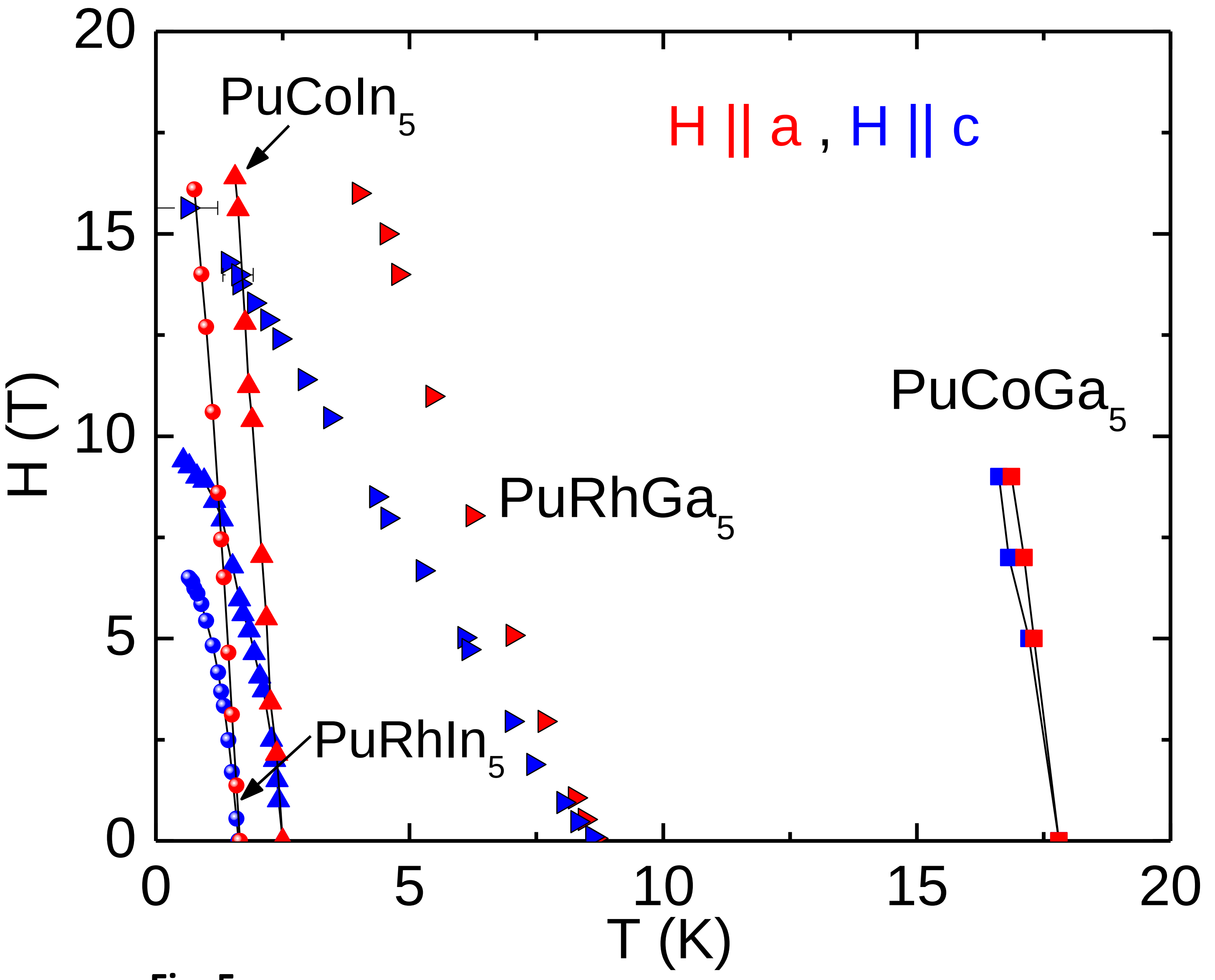

**Fig. 5**

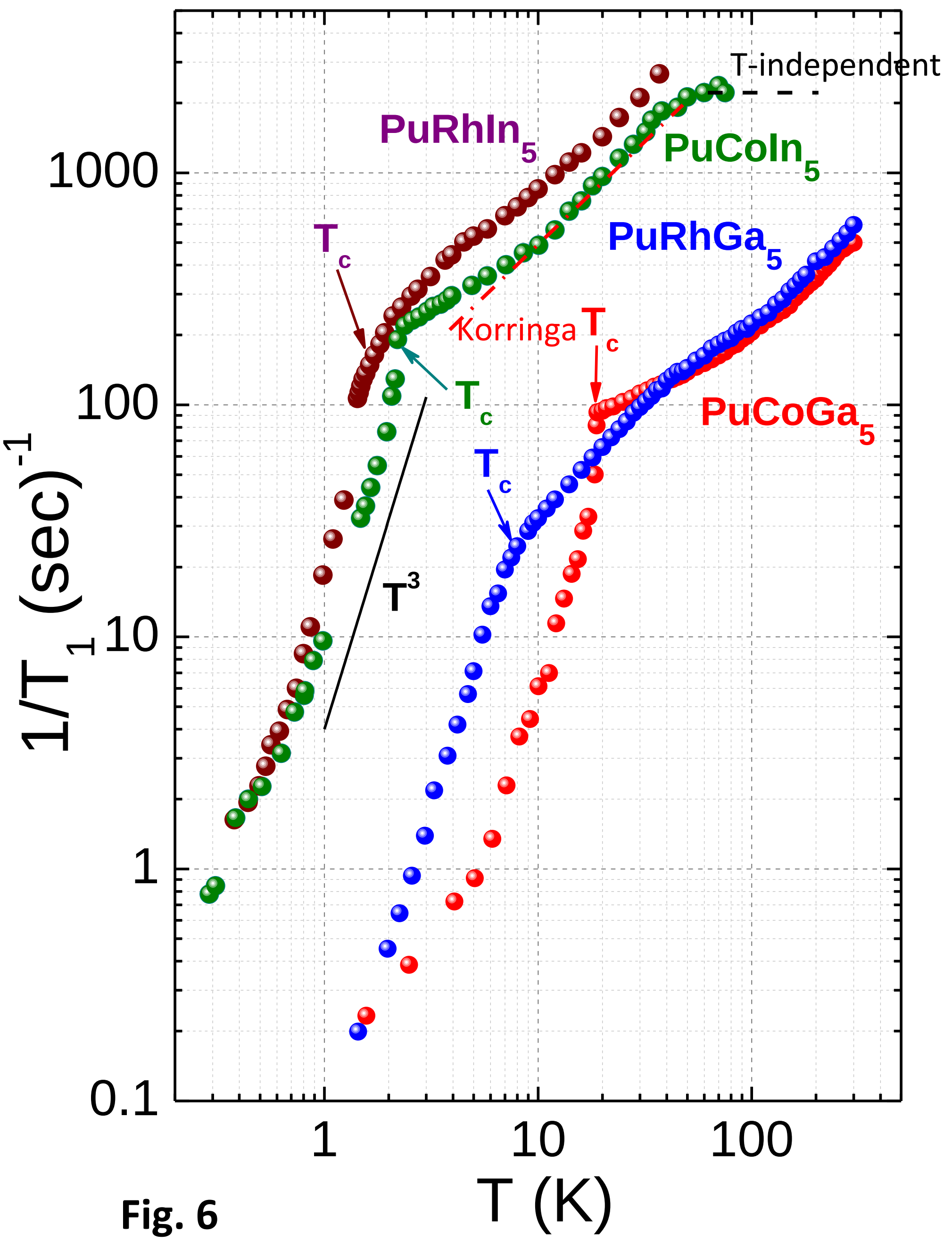

Fig. 6

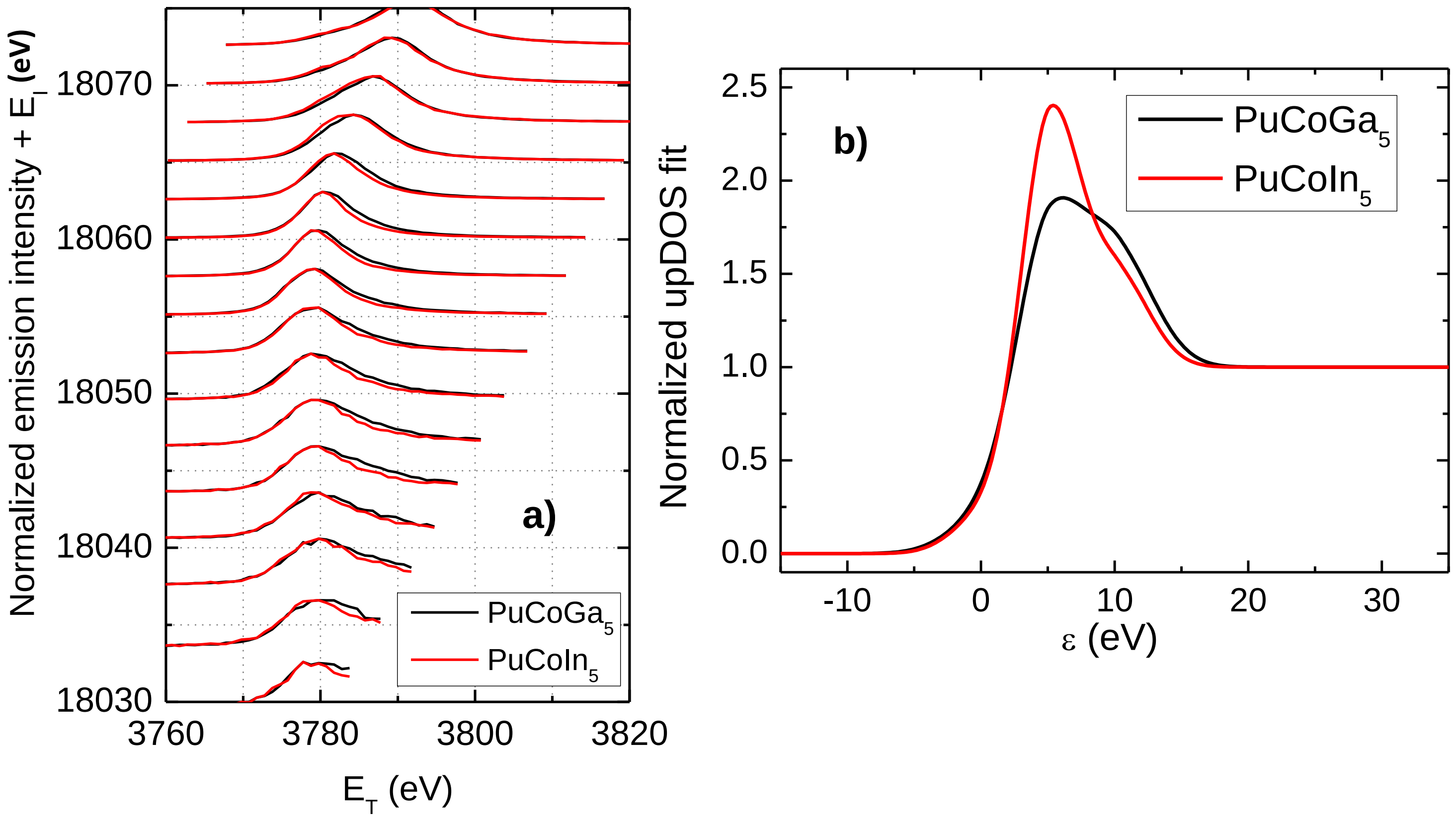

**Fig. 7**

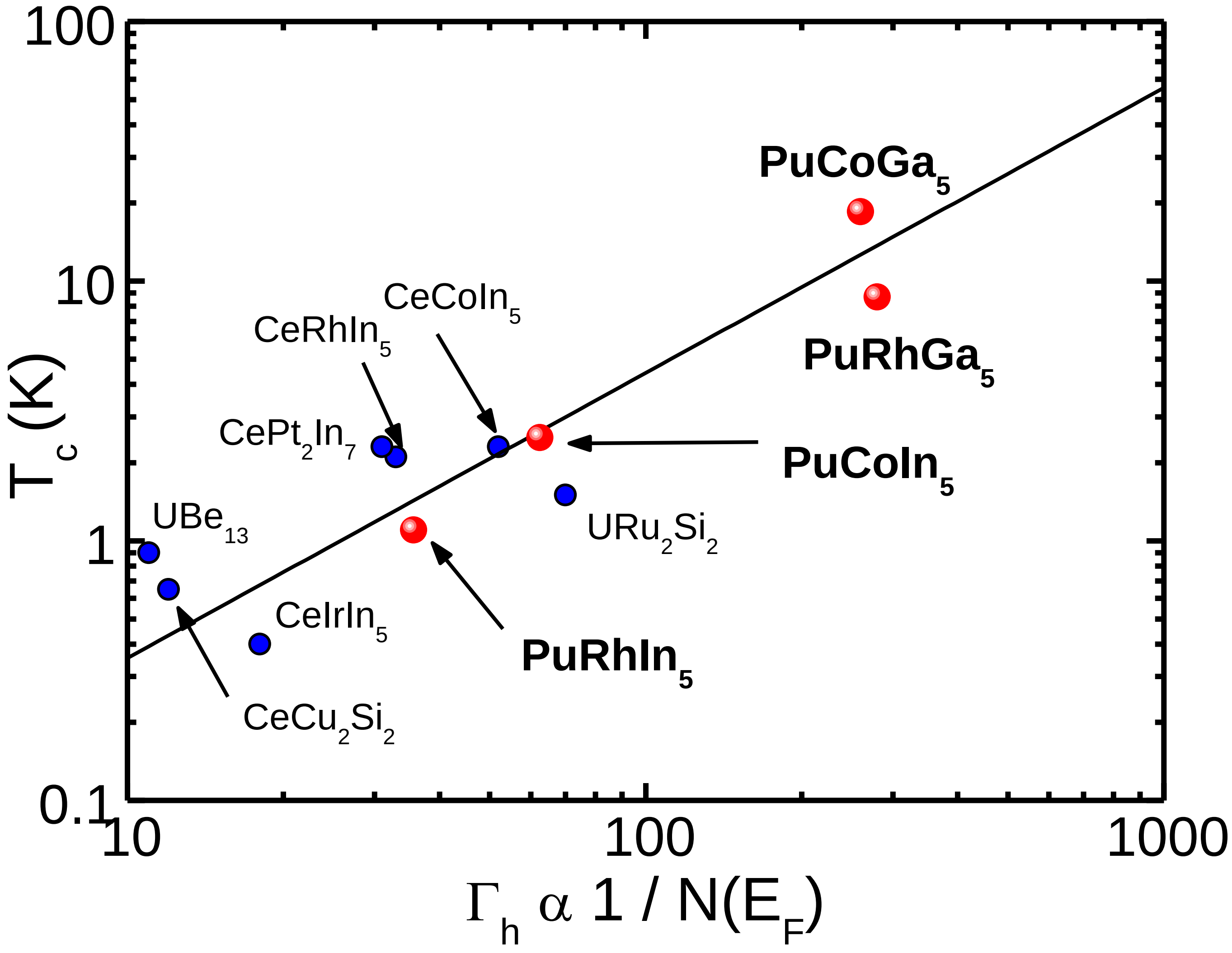

**Fig. 8**